\documentclass[preprint,aps,superscriptaddress]{revtex4-1}
\usepackage{amsmath}
\usepackage{amssymb}
\usepackage{amsfonts}
\usepackage{bm}
\usepackage{amssymb}
\usepackage[dvips]{graphicx}
\usepackage{dcolumn}
\usepackage{txfonts}
\usepackage{makeidx}
\usepackage{color}
\usepackage{mathtools}
\usepackage{threeparttable}
\usepackage[linkcolor=blue,anchorcolor=black,citecolor=blue]{hyperref}
\usepackage{textpos}
\begin{document}

\title{Experimental test of entangled histories}

\begin{textblock}{0.6}[0.5,0.5](11,-0.5)
\fbox{ MIT-CTP-4749}
\end{textblock}

\author{Jordan Cotler}
\affiliation{Stanford Institute for Theoretical Physics, Department of
Physics, Stanford University}

\author{Lu-Ming Duan}\email{lmduan@umich.edu}
\affiliation{Center for Quantum Information, Institute for
  Interdisciplinary Information Sciences, Tsinghua University, Beijing
  100084, China}
\affiliation{Department of Physics, University of Michigan, Ann Arbor, Michigan 48109, USA}

\author{Pan-Yu Hou}
\affiliation{Center for Quantum Information, Institute for
  Interdisciplinary Information Sciences, Tsinghua University, Beijing
  100084, China}

\author{Frank Wilczek}
\affiliation{Center for Theoretical Physics, MIT, Cambridge MA 02139 USA}
\affiliation{Origins Project, Arizona State University, Tempe AZ 25287 USA}

\author{Da Xu}
\affiliation{Center for Quantum Information, Institute for
  Interdisciplinary Information Sciences, Tsinghua University, Beijing
  100084, China}
\affiliation{Department of Physics, Tsinghua University, Beijing
  100084, China}

\author{Zhang-Qi Yin}\email{yinzhangqi@mail.tsinghua.edu.cn}
\affiliation{Center for Quantum Information, Institute for
  Interdisciplinary Information Sciences, Tsinghua University, Beijing
  100084, China}

\author{Chong Zu}
\affiliation{Center for Quantum Information, Institute for
  Interdisciplinary Information Sciences, Tsinghua University, Beijing
  100084, China}

\begin{abstract}
  We propose and demonstrate experimentally a scheme to create entangled
  history states of the Greenberger-Horne-Zeilinger (GHZ) type. In our experiment, the polarization states of a single photon at
  three different times are prepared as a GHZ entangled history state. We define a GHZ functional
  which attains a maximum value $1$ on the ideal GHZ entangled history state and is bounded above by $1/16$
  for any three-time history state lacking tripartite entanglement. We have measured the GHZ functional
  on a state we have prepared experimentally, yielding a value of $0.656\pm 0.005$,  clearly demonstrating the contribution of entangled histories.

\end{abstract}

\maketitle

\section{Introduction}

The traditional ``observables'' of quantum theory are operators in Hilbert space that act at a particular time.  But
many quantities of physical interest, such as the accumulated phase $\exp i \int\limits_1^2 \,  dt \, \vec v \cdot \vec A$ of a particle moving in an electromagnetic potential, or its accumulated proper time, are more naturally expressed in terms of histories.  We may ask: Having performed a measurement of this more general, history-dependent sort of observable, what have we learned?   For conventional observables, the answer is that we learn our system is in a particular subspace of Hilbert space, that is, the eigenspace corresponding to the observable's measured value.   

Recently two of us, building on the work of Griffiths and others \cite{Griffiths1}, have formulated a mathematical framework which extends many of the concepts and procedures ordinarily used in analyzing
{\em states} of quantum systems to their {\em histories} \cite{CW1,CW2,CW2.5}.   Specifically, we have constructed, under very general assumptions about a quantum dynamical system, a Hilbert space of its possible histories.  The inner product reflects probabilities of histories occurring.   

There is a natural definition of observables on the history Hilbert space.  It accommodates the observables we mentioned initially, and new possibilities which might not have been easy to imagine otherwise.   The result of measuring a history observable is a partial reconstruction of ``what happened'' during the evolution of one's system.  Analogously to how measurement of the value of an ordinary observable on a system establishes the location of the state of the system within an eigenspace of the observable, measurement of a history observable on a system's evolution establishes the location of its history within an eigenspace (in history space) of the history observable.   Such eigenspaces often contain {\it entangled\/}  histories.  The defining property of an entangled history is that it cannot be assigned, at each time, to a definite state.  

A particularly interesting sort of entangled history corresponds to a particular state at an initial time, and to another particular state at a final time, and yet can not be assigned to a definite state at intermediate times.   Such an entangled history provides a vivid illustration of the ``many worlds'' picture of quantum mechanics, for it branches into several incompatible trajectories, which later come together.  

Here we describe a detailed protocol for producing an entangled history of that kind.  We have produced histories following that protocol, and measured that they display behavior which cannot be realized by any history which is not entangled.

Our history state is a temporal analogue of the GHZ state, and our measurement strategy was inspired by the GHZ test.   Hence it is appropriate briefly to describe the nature of that test, and its context.

In 1935, Einstein, Podolsky and Rosen (EPR) noted a peculiar consequence of quantum theory, according to which measurement outcomes of distant entangled
particles should be perfectly correlated -- a result they felt to be in tension with relativistic locality (which limits causal influence by the speed of light) \cite{EPR}.
In 1964, John Bell proved that the predictions of quantum theory differ quantitatively from any that can emerge from a large class of 
deterministic (classical) local models \cite{Bell1964}.
Following Bell's suggestion, numerous experiments have been done, and confirmed quantum theory
 \cite{FC1972,Aspect,Hensen2015}.
In 1989,  Greenberger, Horne, and Zeilinger proposed a related test which is slightly more complicated to set up experimentally, but much simpler to interpret and more striking theoretically  \cite{GHZ}.  The GHZ test was then made transparent by Mermin \cite{Mermin1}, brilliantly expounded
by Coleman \cite{Coleman1}, and performed by Pan {\em et al} \cite{Pan2000}.

A particular multipartite entangled state called the GHZ state, which involves at least three spin-$1/2$ degrees of freedom, is essential for the GHZ test.  In our proposal, we deal with the properties of one photon at three different times, rather than three photons at
the same time.  Below, we define a functional
$\mathcal{G}$ which is diagnostic of entangled histories.  We will prove that for product history states $\mathcal{G}$ is bounded above by 0, while for history states lacking tripartite entanglement it is bounded above by $1/16$.
For ideal GHZ entangled histories, the functional $\mathcal{G}$ is
equal to $1$.

Motivated by these ideas, we performed the GHZ test for
entangled histories experimentally. We generated a candidate GHZ entangled history state for a single photon, and  measured its
$\mathcal{G}$ functional.  We measure $\mathcal{G}$  to be  $0.656 \pm 0.005$, which considerably exceeds the bounds mentioned previously. Therefore, the GHZ test for histories clearly demonstrates the existence of entangled histories.

We should add that the structure of our temporal analogue, unlike the original GHZ test, does not preclude its classical modeling.  Indeed, in appropriate limits our setup can be understood on the basis of classical optics (which anticipates key features of quantum theory, i.e. complex waves which interfere, and whose absolute square is physically salient).   Still, it seems to us noteworthy that a simple stochastic classical model for the functional $\mathcal{G}$ has an upper bound of $1/16$.   (See the supplementary materials.)  While that classical model does not map onto our experiment cleanly, its analysis is instructive.

We should also mention that a very interesting, but quite distinct aspect of temporal correlation in quantum theory has been the subject of previous study \cite{Paz, temporalCHSH, fritz1, Waldherr, LGreview}.   These works focus on temporal correlations induced by the usual quantum measurement process, whereas we are primarily concerned with correlations that are intrinsic to the dynamical system.


\section{Constructing an Entangled History State}

In Griffith's theory of ``Consistent Histories" \cite{Griffiths1}, a history state is a sum of tensor-like event products, in the form \begin{equation}
|\Psi) = \hat{P}_{t_{n}}^{i_{n}}
\odot\cdots \odot \hat{P}_{t_{3}}^{i_{3}} \odot\hat{P}_{t_{2}}^{i_{2}}\odot\hat{P}_{t_{1}}^{i_{1}}
\end{equation}
Here the $P_{t_{j}}^{i_{j}}$ are projectors at different times
in temporal order $t_{1}<t_{2}<t_{3}< \cdots < t_{n}$)
where the  index $i$ distinguishes orthogonal projectors within a decomposition of the identity.  $\odot$ is a
typographical variation on the tensor product symbol $\otimes$, which we use when the factors in tensor product
refer to different times.
An inner product on history space is defined by
\begin{equation}
(\Psi | \Phi ) ~=~ \text{Tr}(K^{\dagger}|\Psi) \, K|\Phi))
\end{equation}
where
\begin{equation}
K|\Psi)=\hat{P}_{t_{n}}^{i_{n}}T(t_{n},t_{n-1})\cdots\hat{P}_{t_{3}}^{i_{3}} T(t_{3},t_{2})\hat{P}_{t_{2}}^{i_{2}}T(t_{2},t_{1})\hat{P}_{t_{n}}^{i_{n}}
\end{equation}
Using this positive semi-definite inner product we can define quantum superposition and quantum interference for {\em histories\/} just as we do for quantum states. 
For a more detailed exposition, see \cite{CW1}.

Let us discuss, conceptually, how we might construct the GHZ history state
\begin{equation}
|\text{GHZ}):= \frac{1}{\sqrt{2}} \left( [z^+] \odot [z^+] \odot [z^+]- [z^-] \odot [z^-]\odot [z^-] \right),
\end{equation}
where the notation $\vert \,\cdot\,)$ is used to denote the history state, and where $[z^\pm]:= |z^\pm \rangle \langle z^\pm|$.  Consider a spin-1/2 particle in the state $|x^+\rangle = \frac{1}{\sqrt{2}} (|z^+\rangle + |z^-\rangle)$.  We are going to construct an entangled history state via a post-selection procedure \cite{Aharonov2}.  We introduce three auxiliary qubits $|0\rangle_1 |0\rangle_2 |0\rangle_3 =: |000\rangle$.   At time $t_1$ we perform a CNOT operation between the first auxiliary qubit and the spin-1/2 particle, resulting in
\begin{equation}
\frac{1}{\sqrt{2}} \,|z^+\rangle|000\rangle + \frac{1}{\sqrt{2}} \,|z^-\rangle |100\rangle
\end{equation}
We let this system evolve trivially to time $t_2$.  Then at time $t_2$, we perform a CNOT between the second auxiliary qubit and the spin-1/2 particle, resulting in
\begin{equation}
\frac{1}{\sqrt{2}} \,|z^+\rangle|000\rangle + \frac{1}{\sqrt{2}} \,|z^-\rangle |110\rangle
\end{equation}
The system then evolves trivially to time $t_3$, at which time we perform a CNOT between the third auxiliary qubit and the spin-1/2 particle, giving
\begin{equation}
\frac{1}{\sqrt{2}} \,|z^+\rangle|000\rangle + \frac{1}{\sqrt{2}} \,|z^-\rangle |111\rangle
\end{equation}

If we measure the auxiliary qubits in the $\{|000\rangle, |111\rangle,...\}$ basis, then measuring $|000\rangle$ would indicate that the spin-1/2 particle has been in the history state $[z^+] \odot [z^+] \odot [z^+]$; and if we measure $|111\rangle$, this would indicate that the spin-1/2 particle has been in the history state $[z^-] \odot [z^-] \odot [z^-]$.  However we can also choose to measure the auxiliary qubits in the GHZ basis $\left\{\frac{1}{\sqrt{2}}(|000\rangle \pm |111\rangle),...\right\}$.  Then if we measure $\frac{1}{\sqrt{2}}(|000\rangle - |111\rangle)$, it means that the spin-1/2 particle has been in the history state $[z^+] \odot [z^+] \odot [z^+]$ with amplitude $1/\sqrt{2}$, \textit{and} $[z^-] \odot [z^-] \odot [z^-]$ with amplitude $-1/\sqrt{2}$.  In other words, the particle has been in the entangled history state $\frac{1}{\sqrt{2}} \left( [z^+] \odot [z^+] \odot [z^+]- [z^-] \odot [z^-]\odot [z^-] \right)$.  By changing the basis of the auxiliary qubits, we have \textit{erased} knowledge about the history of the spin-1/2 particle.  As emphasized in \cite{CW3}, selective erasure can be a powerful tool for exploring quantum interference phenomena.

Similar techniques have been proposed in the context of ``multiple-time states" \cite{Aharonov2}.  In this language, we can write the temporal GHZ state as $\frac{1}{\sqrt{2}} \left(\langle z^+|\,|z^+\rangle \langle z^+|\,|z^+\rangle \langle z^+|\,|z^+\rangle - \langle z^-|\,|z^-\rangle \langle z^-|\,|z^-\rangle \langle z^-|\,|z^-\rangle \right)$.  The interpretation of temporal entanglement has been a subject of much debate \cite{Aharonov2}, \cite{Dana}-\cite{Megidish}.   The framework proposed in \cite{CW1, CW2, CW2.5}, grounded in the consistent histories approach of Griffiths \cite{Griffiths1}, seems to us clear and unambiguous.

\section{Temporal GHZ Test}
In this section we will discuss how to perform a GHZ test for entangled histories.  Consider the operators
\begin{equation}
\label{ops1}
\sigma_x \odot \sigma_y \odot \sigma_y\,\,,\qquad \sigma_y \odot \sigma_x \odot \sigma_y\,\,,\qquad \sigma_y \odot \sigma_y \odot \sigma_x\,\,,\qquad\sigma_x \odot \sigma_x \odot \sigma_x
\end{equation}
on a three-time history space of a single spin-1/2 particle with trivial time evolution.  The expectation values of $|\text{GHZ})$ with the history state operators corresponding to the four operators in Equation \ref{ops1} are $1$, $1$, $1$ and $-1$, respectively.  The product of these four expectation values is $-1$.  We represent this procedure of computing the product of the expectation values by the functional $\mathcal{G}$ which satisfies
\begin{equation}
\label{Gfunctional1}
\mathcal{G}[|\Psi)] =  -\langle \sigma_x \odot \sigma_x \odot \sigma_x \rangle \langle \sigma_y \odot \sigma_y \odot \sigma_x \rangle \langle \sigma_y \odot \sigma_x \odot \sigma_y \rangle \langle \sigma_x \odot \sigma_y \odot \sigma_y \rangle
\end{equation}
where $|\Psi)$ is a normalized history state.  For the ideal GHZ history state, we have
\begin{equation}
\label{Gfunctional2}
\mathcal{G}[|\text{GHZ})] = 1
\end{equation}

We can write a history state with two-time entanglement as
 $|\psi) \odot \left[\begin{array}{c c}
	\cos^2(\theta) & \cos(\theta) \sin(\theta) \,e^{-i\phi}\\
	\cos(\theta) \sin(\theta)\,e^{i\phi} & \sin^2(\theta)
	\end{array} \right]$, where $|\psi)$ is arbitrary two-time entangled history states. Other history states with two-time entanglement take the same form, up to permutations of the tensor product components.  Since the $\mathcal{G}$ functional is not sensitive to such permutations, it suffices to consider a single ordering.  We have
proved that (supplementary materials)
	\begin{align}
	&\mathcal{G}\left(|\psi) \odot \left[\begin{array}{c c}
	\cos^2(\theta) & \cos(\theta) \sin(\theta) \,e^{-i\phi} \\
	\cos(\theta) \sin(\theta)\,e^{i\phi} & \sin^2(\theta)
	\end{array} \right] \right) \nonumber \\
	& \qquad \qquad \qquad \qquad \qquad =  - \frac{1}{4}\sin^{4}(2\theta)\sin^{2}(2\phi)\langle\sigma_{x}\odot\sigma_{x}\rangle
\langle\sigma_{y}\odot\sigma_{y}\rangle\langle\sigma_{x}\odot\sigma_{y}\rangle\langle\sigma_{y}\odot\sigma_{x}\rangle \leq \frac{1}{16}
	\end{align}
And for a generic separable history state
\begin{equation}
| \psi_{\rm pure} ) ~\equiv~ P(\theta_1, \phi_1) \odot P(\theta_2, \phi_2) \odot P(\theta_3, \phi_3)
\end{equation}
where $P(\theta, \phi) = \left[\begin{array}{c c}
	\cos^2(\theta) & \cos(\theta) \sin(\theta) \,e^{-i\phi}\\
	\cos(\theta) \sin(\theta)\,e^{i\phi} & \sin^2(\theta)
	\end{array} \right]$\,,
we have
\begin{equation}
\mathcal{G}[|\psi_{\rm pure} )] ~=~ -\frac{1}{64} \, \sin^4(2\theta_1) \sin^4(2\theta_2) \sin^4(2\theta_3) \sin^2(2\phi_1) \sin^2(2\phi_2) \sin^2(2\phi_3) \leq 0
\end{equation}


Our goal is to construct an approximation to
the history state $|\text{GHZ})$ experimentally, and to show that for our constructed state $\mathcal{G}[|\text{GHZ})] \gg  1/16$, thus demonstrating a high degree of temporal
entanglement.  (In fact the $\mathcal{G}$ functional even distinguishes a specific form of tripartite entanglement.  For the W entangled history state, $|W)= \frac{1}{\sqrt{3}} \left( [z^-] \odot [z^+] \odot [z^+]+ [z^+] \odot [z^-]\odot [z^+] + [z^+] \odot [z^+] \odot [z^-] \right)$, the $\mathcal{G}$ functional vanishes.)  This $\tau$GHZ test is much simpler than the generalized temporal Bell test in \cite{CW2}, and
requires many fewer measurements.

\newpage
\section{Experimental results}

We have phrased our discussion in the language appropriate to the spin states of a spin-$\frac{1}{2}$
particle. As is well known, we can use the same two-dimensional, complex state space to describe the
polarization states of a photon.  In that context, it is known as the Poincar\'{e} sphere.
We can adapt standard optical tools and techniques to create a temporal GHZ state for a photon
and measure the correlations that appear in the GHZ functional. The predicted correlations, as we have
seen, provide quantitative evidence for the contribution of highly entangled histories.

Before proceeding, we provide a ``dictionary" between the complex state space of a spin-1/2 particle and the polarization state of a single photon.  We make the identifications
\begin{align*}
|z^+\rangle &\,\,\longleftrightarrow\,\, |H\rangle \\
|z^-\rangle &\,\,\longleftrightarrow\,\, |V\rangle \\
|x^+\rangle &\,\,\longleftrightarrow\,\, |D\rangle \\
|x^-\rangle &\,\,\longleftrightarrow\,\, |A\rangle \\
|y^+\rangle &\,\,\longleftrightarrow\,\, |R\rangle \\
|y^-\rangle &\,\,\longleftrightarrow\,\, |L\rangle
\end{align*}
where ``H" stands for ``horizontally" polarized light, ``V" stands for ``vertically" polarized light, ``D" stands for ``diagonally" polarized light, ``A" stands for ``anti-diagonally" polarized light, ``R" stands for ``right-circularly" polarized light, and ``L" stands for ``left-circularly" polarized light.  We have the standard relations
\begin{align}
|D\rangle &= \frac{1}{\sqrt{2}} \left(|H\rangle + |V\rangle \right) \\
|A\rangle &= \frac{1}{\sqrt{2}} \left(|H\rangle - |V\rangle \right) \\
|R\rangle &= \frac{1}{\sqrt{2}} \left(|H\rangle + i|V\rangle \right) \\
|L\rangle &= \frac{1}{\sqrt{2}} \left(|H\rangle - i|V\rangle \right)
\end{align}

To  implement the GHZ test for entangled histories experimentally, we prepare a single photon through
spontaneous parametric down conversion (SPDC) shown in Figure 2. The SPDC process generates photon pairs
with perpendicular polarizations, which are then separated by a polarizing
beamsplitter (PBS). Through detection of the reflected photon by an avalanche photodiode single photon detector (D1), we
get a  single photon source on the other outport (Fiber coupler 3). We prepare this (approximate) single photon as a
diagonal polarization state $[D]$ with a fiber-based polarization controller (PC) and a polarizer, and then send it into a balanced Mach-Zehnder interferometer (MZI), each arm of which supports a sequence of PBSs and wave-plate sets (WP). The incoming photon in the $[D]$ state is initially split, by PBS0, into horizontal (H) and vertical (V) components of equal amplitude, traveling along the two arms. WP2 and PBS2 divide the photon in the lower arm again, removing one polarization direction, while the other continues along the arm.  We might, for example, remove the $[L]$ polarization, while allowing $[R]$ to continue propagating. WP4 then rotates the propagating photon back to $[H]$ direction.  Two more operations (PBS4, WP6 and PBS6, WP8) of the same type take place, until the surviving photon reaches PBS7. The surviving photon will have been in the history state $[H] \odot [H] \odot [H]$, and sampled by the observable $[R] \odot * \odot *$, where the wild cards reflect our choices of polarization for PBS4 and PBS6. A completely parallel analysis applies to the other arm. Finally, the surviving components recombine coherently at PBS7, another polarizing beam splitter, and emerge in a direction that enforces a relative minus sign between the contributions from $[H] \odot [H] \odot [H]$ and $[V] \odot [V] \odot [V]$. By post-selecting on the events that trigger D2, and varying the wave-plate sets appropriately, we can measure the GHZ functional for an entangled history.  We must measure the expectation values $\langle \sigma_x \odot \sigma_x \odot \sigma_x \rangle$, $\langle \sigma_y \odot \sigma_y \odot \sigma_x \rangle$, $\langle \sigma_y \odot \sigma_x \odot \sigma_y \rangle$ and $\langle \sigma_x \odot \sigma_y \odot \sigma_y \rangle$ with respect to the GHZ history state, and then multiply all of the expectation values together.

In this experiment we often access polarization properties at definite times, through PBS$1-6$. Alternatively, in the case where the photon transmits at all of the 6 PBSs and triggers D2, we access a multi-time observable. Such multi-time observables represent, in Griffiths' terminology \cite{Griffiths1},  ``contextual'' properties. We can form a family based on those complementary single-time and multi-time properties. The character of the history state $\frac{1}{\sqrt 2}([z^+] \odot [z^+] \odot [z^+] - [z^-] \odot [z^-] \odot [z^-])$ emerges clearly only when we measure multi-time observables, which in turn we access when certain other events fail to occur.

The experimental procedure is divided into 32 trials, which each have separate settings for the polarizing beam splitters and wave plates. Each trial consists of preparing the settings of the apparatus and recording the number of photons received by detector D2 (denoted by $\text{Counts}(\text{D2})$). \newpage \noindent We define
\begin{align}
&x_1 = D\,\,, \qquad x_2 = A\,\,, \qquad x_3 = R\,\,, \qquad x_4 = L \nonumber \\
&\overline{x}_1 = A\,\,, \qquad \overline{x}_2 = D\,\,, \qquad \overline{x}_3 = L\,\,, \qquad \overline{x}_4 = R \nonumber
\end{align}
Let PBS$(\alpha,\overline{\alpha})$ denote that the PBS transmits the photon in the $|\alpha\rangle$ polarization and reflects the orthogonal component $|\overline{\alpha}\rangle$, while WP$(\beta, \gamma)$ denote that the WP is set up so that the incoming photon in the $|\beta\rangle$ polarization is transformed into the $|\gamma\rangle$ polarization. Then for a given trial, the settings have the form
\begin{align*}
\{\,&\text{PBS1}(x_i,\overline{x}_i),\,\text{WP3}(x_i,\text{H}), \,\text{PBS2}(x_i,\overline{x}_i),\,\text{WP4}(x_i,\text{V}),\\&\text{PBS3}(x_j,\overline{x}_j),\,\text{WP5}(x_j,\text{H}), \, \text{PBS4}(x_j,\overline{x}_j),\,\text{WP6}(x_j,\text{V}), \\
&\text{PBS5}(x_k,\overline{x}_k),\,\text{WP7}(x_k,\text{H}), \,\text{PBS6}(x_k,\overline{x}_k),\,\text{WP8}(x_k,\text{V})\,\}
\end{align*}
with fixed values of $i,j,k = 1,2,3$.

We define
\begin{align*}
C_{i,j,k} &:= \frac{\text{Counts}_{i,j,k}(\text{D2})}{\text{Counts}(\text{total})}
\end{align*}
where ${\text{Counts}(\text{total})}$ denotes the overall number of photons sent into the MZI, which is a constant for different trials in our experiment.  By collecting data from all of the trials we can evaluate
\begin{align}
\label{expect1}
\langle \sigma_x \odot \sigma_x \odot \sigma_x \rangle = &\frac{C_{1,1,1} - C_{1,1,2} - C_{1,2,1} + C_{1,2,2} - C_{2,1,1} + C_{2,1,2} + C_{2,2,1} - C_{2,2,2}}{C_{1,1,1} + C_{1,1,2} + C_{1,2,1} + C_{1,2,2} + C_{2,1,1} + C_{2,1,2} + C_{2,2,1} + C_{2,2,2}}
\end{align}
\begin{align}
\label{expect2}
\langle \sigma_y \odot \sigma_y \odot \sigma_x \rangle = &\frac{C_{3,3,1} - C_{3,3,2} - C_{3,4,1} + C_{3,4,2} - C_{4,3,1} + C_{4,3,2} + C_{4,4,1} - C_{4,4,2}}{C_{3,3,1} + C_{3,3,2} + C_{3,4,1} + C_{3,4,2} + C_{4,3,1} + C_{4,3,2} + C_{4,4,1} + C_{4,4,2}}
\end{align}
\begin{align}
\label{expect3}
\langle \sigma_y \odot \sigma_x \odot \sigma_y \rangle = &\frac{C_{3,1,3} - C_{3,1,4} - C_{3,2,3} + C_{3,2,4} - C_{4,1,3} + C_{4,1,4} + C_{4,2,3} - C_{4,2,4}}{C_{3,1,3} + C_{3,1,4} + C_{3,2,3} + C_{3,2,4} + C_{4,1,3} + C_{4,1,4} + C_{4,2,3} + C_{4,2,4}}
\end{align}
\begin{align}
\label{expect4}
\langle \sigma_x \odot \sigma_y \odot \sigma_y \rangle = &\frac{C_{1,3,3} - C_{1,3,4} - C_{1,4,3} + C_{1,4,4} - C_{2,3,3} + C_{2,3,4} + C_{2,4,3} - C_{2,4,4}}{C_{1,3,3} + C_{1,3,4} + C_{1,4,3} + C_{1,4,4} + C_{2,3,3} + C_{2,3,4} + C_{2,4,3} + C_{2,4,4}}
\end{align}
and finally to compute
\begin{equation*}
\mathcal{G}[|\Psi)] = - \langle \sigma_x \odot \sigma_x \odot \sigma_x \rangle \langle \sigma_y \odot \sigma_y \odot \sigma_x \rangle \langle \sigma_y \odot \sigma_x \odot \sigma_y \rangle \langle \sigma_x \odot \sigma_y \odot \sigma_y \rangle
\end{equation*}
by taking the product.

\begin{widetext}
\begin{figure}[]
  \includegraphics[width=100mm]{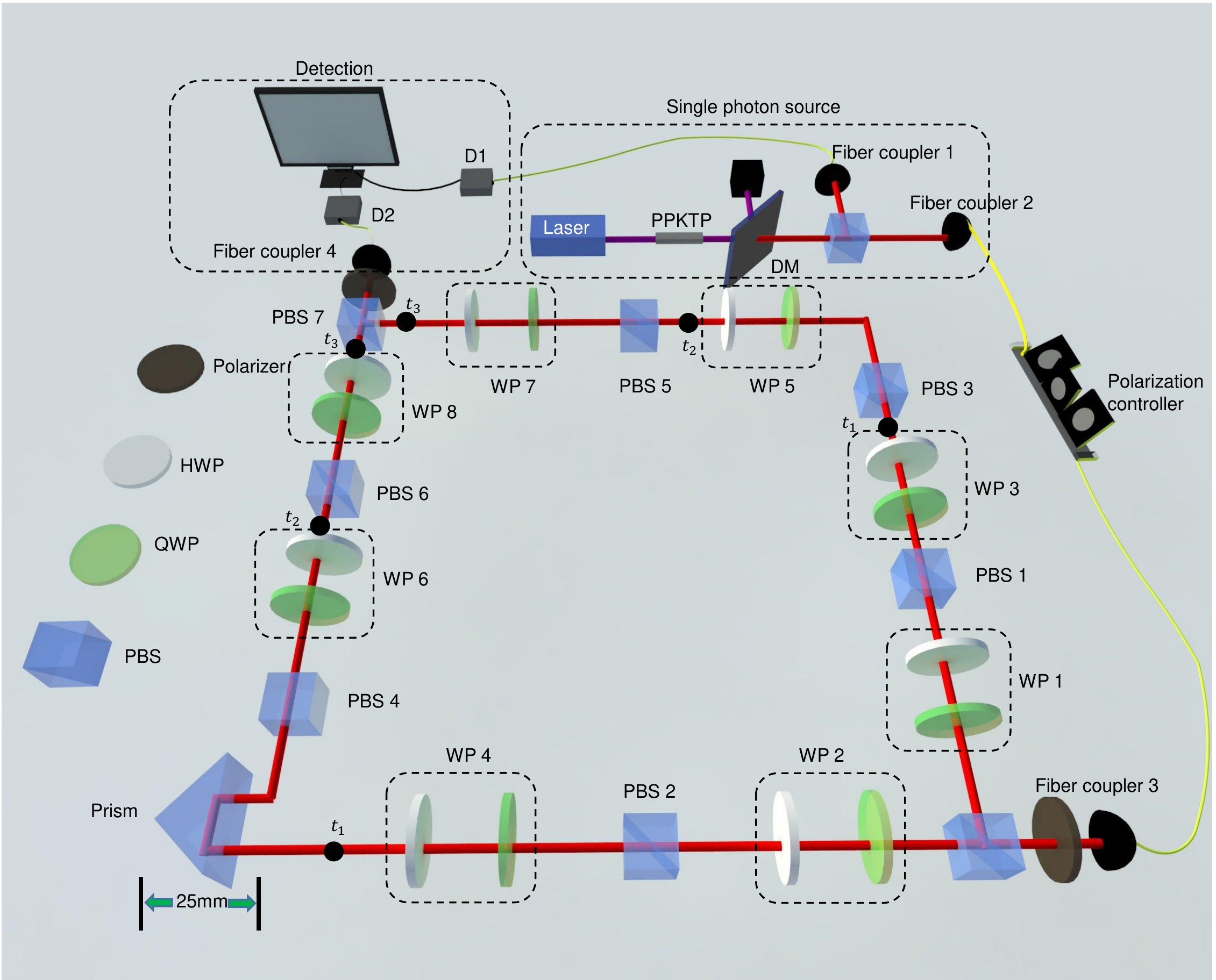}
  \caption{Illustration of the experimental setup. A continuous-wave diode laser around $404$ nm in wavelength, after a band-pass filter centered at $404$ nm with $3$ nm bandwidth, is focused on a type II PPKTP crystal to generate correlated photon pairs of $808$ nm wavelength with perpendicular polarizations through spontaneous parametric down conversion.  A dichroic mirror (DM) is used to filter out the pump beam. The photon pairs are split by a polarizing beam splitter (PBS), and then coupled into single mode fibers. With the registration of a photon count at a fiber-based single photon detector D1, we get a heralded single photon source in the other fiber outport (fiber coupler 3). The polarization of the heralded photon is set to the state $|D\rangle = (|H\rangle+|V\rangle)/\sqrt{2}$ with a fiber-based polarization controller (PC). After filtering by a polarizer oriented at $45^{\circ}$, the photon is sent into a Mach-Zehnder interferometer (MZI) with two arms of equal lengths. In the MZI, a set of half-wave plates (HWP), quarter-wave plates (QWP), and PBSs are applied at $3$ different times (denoted as $t_1$, $t_2$ and $t_3$ in the figure) to perform the projective measurements and polarization recovery operations for the $\tau$GHZ test. All of the wave-plates are mounted on motorized precision rotation mounts 
that are automatically controlled by a computer. A prism, positioned on a one-axis motorized translation stage
, is used to precisely adjust the length of one arm so that the two spatial modes in the MZI coherently interfere with each other at the PBS7 before readout by detector D2. We register the two-fold coincidence counts between D1 and D2 with a $5$ ns window through a home-made Field-Programmable Gate Array (FPGA) board. The GHZ test is repeated with $32$ trials, each with different angles of the wave plates (see supplementary materials). To guarantee that the phase of the MZI is stable during the measurement, we monitor the count rate $C_{ref}$ with a fixed wave-plate setting before and after each trial.}
\end{figure}
\end{widetext}

\begin{figure}[]
  \includegraphics[width=130mm]{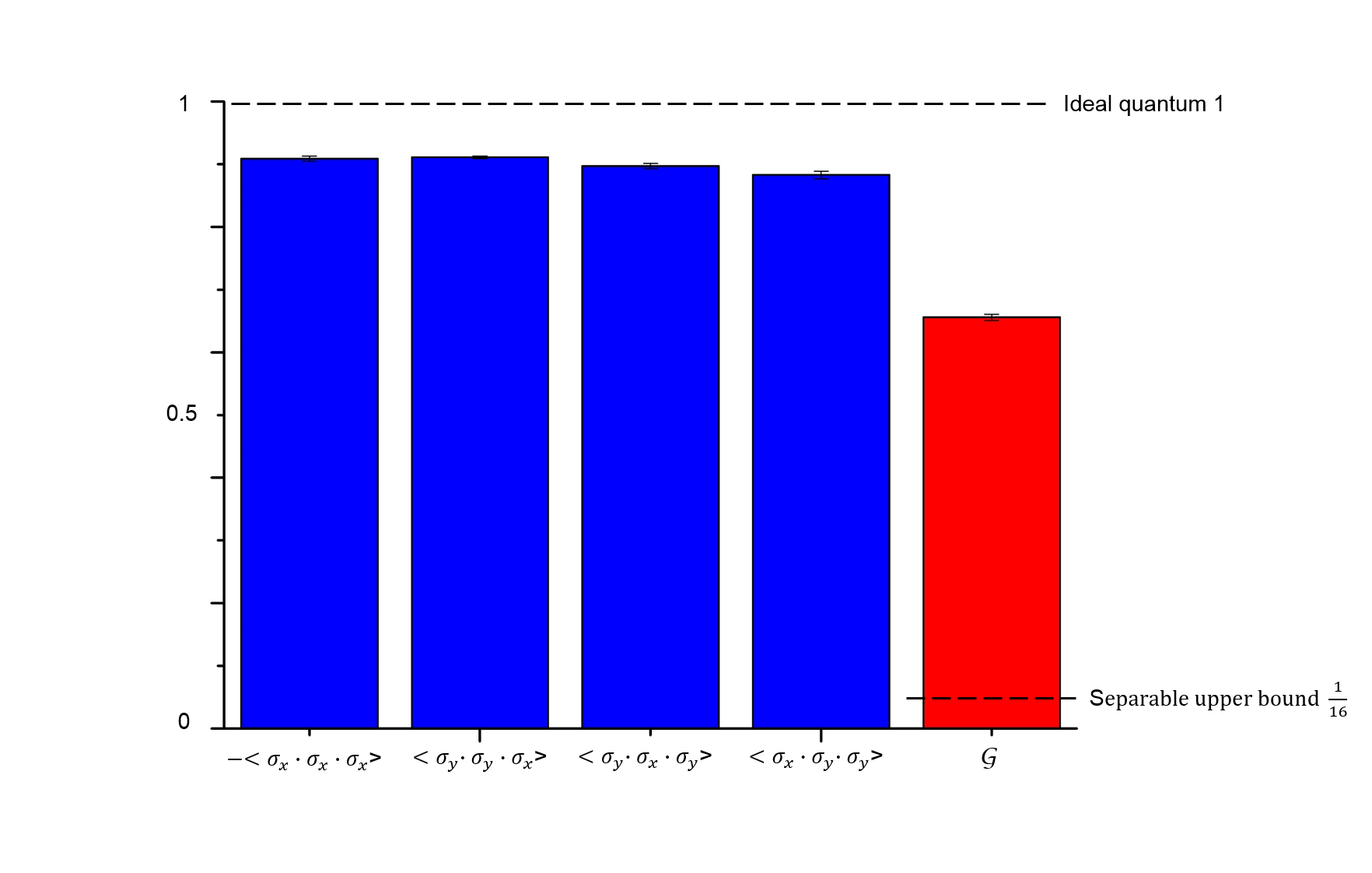}
  \caption{We show the measurement results $(0.909(4), 0.911(2), 0.897(4), 0.883(6))$ corresponding to four different bases and the measured functional $\mathcal{G}= 0.656(5)$ which is clearly larger than $1/16$. The error bar is the standard deviation assuming a Poissonian distribution for the photon counts of the detectors.}\label{fig:data}
\end{figure}

The four measured correlations are shown in Fig. \ref{fig:data}.  The computed value of $\mathcal{G}$ is $0.656\pm0.005$, where the error takes into account the statistics of detector photocounting.

\newpage
\section{Conclusion}
We have performed an experiment to create and validate an entangled history
state for a single photon. The experiment allows us to superpose radically different versions of
the system's history, and to probe its entangled structure.  The experimental results
violate inequalities implied by separability by a large margin.

It may seem startling that any version of the GHZ phenomenon, which in many ways epitomizes the peculiar characteristics of quantum theory, can be reproduced using, in essence, classical optics.   On the other hand, we may recall that Dirac's magisterial text on quantum theory begins with a long discussion of experiments with polarized light \cite{Dirac1}.  Indeed, the wave theory of light already supports the principle of superposition, the calculation of intensities through squares of amplitudes, and interference phenomena -- i.e., the central aspects of quantum-mechanical wave functions.  In our context, the central innovation of quantum theory is not to change the rules of wave theory, but to allow an alternative (particle) interpretation of its results.  By following out that particle interpretation fully we discover that it brings in new ideas, such as the temporal entanglement of histories.

\section*{Acknowledgements}

\noindent JC and FW are grateful to Sandu Popescu and Michael Walter for discussions and valuable insights.  JC is supported by the Fannie and John Hertz Foundation and the Stanford Graduate Fellowship program.  FW's work is supported by the U.S. Department of Energy under grant Contract Number DE-SC0012567. LMD acknowledges support from the IARPA MUSIQC program, the AFOSR and the ARO MURI program. ZQY is supported by the National Natural Science Foundation of China Grant 61435007.

\newpage
\section*{Supplementary Materials}

\subsection{A classical calculation}

Here we analyze a class of classical systems which have qualitative similarities to the $\tau$GHZ test.  
The assumptions for our calculations are as follows. Suppose that we have three observers $A$, $B$, $C$ who measure the value of $\sigma_x$ or $\sigma_y$ at three different times $t_1$, $t_2$ and $t_3$. The measurement basis (the $x$-basis or the $y$-basis) is chosen randomly at each time.  Communication among $A$, $B$ and $C$ is forbidden for $t_1 \leq t \leq t_3$. Then at a later time $t_4$ (after time $t_3$), $A$, $B$ and $C$ exchange information on their choice of basis and measurement
results.  When their basis choices correspond to the $\mathcal{G}$ functional choices (e.g. $\sigma_x$
for $t_1$,
$\sigma_x$ for $t_2$,
and $\sigma_x$
for $t_3$), and only then, the measured results are recorded.

The observable values at the three different times
are allowed to be correlated.   Thus we introduce a master joint total distribution $\chi(Q^1_x ,Q^1_y ,Q^2_x ,Q^2_y ,Q^3_x ,Q^3_y )$, and calculate
\begin{equation}
\mathcal{G}[\chi] ~=~ -( Q^1_x Q^2_x Q^3_x )_\chi ( Q^1_x Q^2_y Q^3_y )_\chi ( Q^1_y Q^2_x Q^3_y )_\chi ( Q^1_y Q^2_y Q^3_x )_\chi
\end{equation}
in an evident notation.

There are eight possible assignments of the products of these variables consistent with the classical constraint that the product of their products is unity.  Let us define their probabilities with respect to the distribution $\chi$ as follows:
\begin{eqnarray}
{\rm Prob} (Q^1_x Q^2_x Q^3_x = +1,  Q^1_x Q^2_y Q^3_y = +1, Q^1_y Q^2_x Q^3_y = +1, Q^1_y Q^2_y Q^3_x =+1) ~&\equiv&~ p_1 \nonumber \\
{\rm Prob} (Q^1_x Q^2_x Q^3_x= +1,  Q^1_x Q^2_y Q^3_y = -1, Q^1_y Q^2_x Q^3_y = -1, Q^1_y Q^2_y Q^3_x =+1) ~&\equiv&~ p_2 \nonumber \\
{\rm Prob} (Q^1_x Q^2_x Q^3_x = +1,  Q^1_x Q^2_y Q^3_y = -1, Q^1_y Q^2_x Q^3_y = +1, Q^1_y Q^2_y Q^3_x =-1) ~&\equiv&~ p_3 \nonumber \\
{\rm Prob} (Q^1_x Q^2_x Q^3_x = +1,  Q^1_x Q^2_y Q^3_y = +1, Q^1_y Q^2_x Q^3_y = -1, Q^1_y Q^2_y Q^3_x =-1) ~&\equiv&~ p_4 \nonumber \\
{\rm Prob} (Q^1_x Q^2_x Q^3_x = -1,  Q^1_x Q^2_y Q^3_y = -1, Q^1_y Q^2_x Q^3_y = -1, Q^1_y Q^2_y Q^3_x =-1) ~&\equiv&~ p_5 \nonumber \\
{\rm Prob} (Q^1_x Q^2_x Q^3_x = -1,  Q^1_x Q^2_y Q^3_y = -1, Q^1_y Q^2_x Q^3_y = +1, Q^1_y Q^2_y Q^3_x =+1) ~&\equiv&~ p_6 \nonumber \\
{\rm Prob} (Q^1_x Q^2_x Q^3_x = -1,  Q^1_x Q^2_y Q^3_y = +1, Q^1_y Q^2_x Q^3_y = -1, Q^1_y Q^2_y Q^3_x =+1) ~&\equiv&~ p_7 \nonumber \\
{\rm Prob} (Q^1_x Q^2_x Q^3_x = -1,  Q^1_x Q^2_y Q^3_y= +1, Q^1_y Q^2_x Q^3_y = +1, Q^1_y Q^2_y Q^3_x = -1) ~&\equiv&~ p_8 \nonumber \\
{} &{}&
\end{eqnarray}
Then for the GHZ functional we have
\begin{eqnarray}
\mathcal{G}[\chi] ~&=&~ -(p_1 + p_2 + p_3 + p_4 - p_5 - p_6 - p_7 -p_8)(p_1 + p_2 - p_3 -p_4 -p_5 + p_6 + p_7 - p_8)  \nonumber \\
~&{}&~ (p_1 - p_2 +p_3 -p_4 -p_5 + p_6 -p_7 + p_8)(p_1 -p_2 - p_3 +p_4 -p_5 -p_6 + p_7 + p_8) \nonumber \\
~&{}&~
\end{eqnarray}
Maximizing this over probability distributions, we find that it is bounded below by
\begin{equation}\label{classicalBound}
\mathcal{G}[\chi] ~\leq~ \frac{1}{16}
\end{equation}
The maximum is assumed at $(p_1, p_2, p_3, p_4, p_5, p_6, p_7, p_8) = (\frac{1}{4}, \frac{1}{4}, \frac{1}{4}, 0, 0, \frac{1}{4}, 0,0)$ and at several other points.  It is intriguing that Eqn.\,(\ref{classicalBound}) is the same bound satisfied by history states lacking tripartite entanglement.

\subsection{Partially separable history state}
We consider a two-time entangled history with an attached separable history state.  Such a state has the form $|\psi) \odot  \left[\begin{array}{c c}
	\cos^2(\theta) & \cos(\theta) \sin(\theta) \,e^{-i\phi} \\
	\cos(\theta) \sin(\theta)\,e^{i\phi} & \sin^2(\theta)
	\end{array} \right]$, where $|\psi)$ is an arbitrary two-time entangled history states.  As mentioned in the paper, other history states with two-time entanglement take the same form, up to permutations of the tensor product components.  It suffices to consider a single ordering since the $\mathcal{G}$ functional is not sensitive to such permutations.
The above history state gives
\begin{equation}
\label{eq:partial}
\mathcal{G}= - \frac{1}{4}\sin^{4}(2\theta)\sin^{2}(2\phi) \langle\sigma_{X}\odot\sigma_{X}\rangle \langle \sigma_{X}\odot\sigma_{Y}\rangle
\langle\sigma_{Y}\odot\sigma_{X}\rangle \langle \sigma_{Y}\odot\sigma_{Y}\rangle
\end{equation}
The term $\frac{1}{4}(\sigma_{X}\odot\sigma_{X})( \sigma_{X}\odot\sigma_{Y})(\sigma_{Y}\odot\sigma_{X})(\sigma_{Y}\odot\sigma_{Y})$ can be written by expanding in a basis of temporal Bell states.
Let us write
\begin{eqnarray}
|\psi)=a|\phi^{+})+b|\phi^{-})+c|\psi^{+})+d|\psi^{-})
\end{eqnarray}
where the basis vectors are the four temporal Bell states, and  $|a|^{2}+|b|^{2}+|c|^{2}+|d|^{2}=1$. For the two-time entangled history state $|\psi)$, we get
\begin{equation}
\mathcal{G}[|\psi)]= -((|c|^{2}-|d|^{2})^2-(|a|^{2}-|b|^{2})^2)((a^*b-b^*a)^{2}-(c^*d-d^*c)^{2})
\end{equation}
By adding extra phase parameters $\phi_{ab}$ and $\phi_{cd}$ we can write Eq. \eqref{eq:partial} as
\begin{eqnarray}
\mathcal{G}=-\sin^{4}(2\theta)\sin^{2}(2\phi)((|c|^{2}-|d|^{2})^2-(|a|^{2}-|b|^{2})^2)((|a||b|\sin(\phi_{ab}))^{2}-(|c||d|\sin(\phi_{cd}))^{2})
\end{eqnarray}
Optimizing over $|a|, |b|, |c|, |d|, \phi_{ab}$ and $\phi_{cd}$ subject to the appropriate constraints, we find that for the class of history states under consideration, the maximum value for $\mathcal{G}$ is exactly $1/16$.

\newpage
\subsection{Wave-plates parameters for experiment}

\begin{table}[!hbp]
\resizebox{\textwidth}{10 cm}{%
\begin{tabular}{|c|c|c|c|c|c|c|c|c|c|c|c|c|c|c|c|c|}
\hline
&    QWP1 &     HWP1&   QWP2&   HWP2&    QWP3&   HWP3&   QWP4&   HWP4&   QWP5&   HWP5&   QWP6&   HWP6&   QWP7&   HWP7&   QWP8&   HWP8\\
\hline
  Ref &  0	&  45	&  0	&  0	&  0	&  0	&  0	&  45		&  0	&  0	&  0	&  0	&  0	&  0	&  0	&  0\\
\hline
	  Trial1.1	&    135	&  67.5	&  90	&  112.5	&  90	&  112.5	&  90	&  45		&  135	&  67.5	&  90	&  67.5	&  90	&  67.5	&  90	&  90\\
\hline
	  Trial1.2	&  45	&  112.5	&  90	&  112.5	&  90	&  112.5	&  90	&  45		&  45	&  112.5	&  90	&  67.5	&  90	&  67.5	&  90	&  90\\
\hline
	  Trial1.3	&  135	&  67.5	&  90	&  67.5	&  90	&  112.5	&  90	&  45		&  135	&  67.5	&  90	&  22.5	&  90	&  67.5	&  90	&  90\\
\hline
	  Trial1.4	&  45	&  112.5	&  90	&  67.5	&  90	&  112.5	&  90	&  45		&  45	&  112.5	&  90	&  22.5	&  90	&  67.5	&  90	&  90\\
\hline
	  Trial1.5	&  135	&  67.5	&  90	&  112.5	&  90	&  67.5	&  90	&  45		&  135	&  67.5	&  90	&  67.5	&  90	&  22.5	&  90	&  90\\
\hline
	  Trial1.6	&  45	&  112.5	&  90	&  112.5	&  90	&  67.5	&  90	&  45		&  45	&  112.5	&  90	&  67.5	&  90	&  22.5	&  90	&  90\\
\hline
	  Trial1.7	&  135	&  67.5	&  90	&  67.5	&  90	&  67.5	&  90	&  45		&  135	&  67.5	&  90	&  22.5	&  90	&  22.5	&  90	&  90\\
\hline
	  Trial1.8	&  45	&  112.5	&  90	&  67.5	&  90	&  67.5	&  90	&  45		&  45	&  112.5	&  90	&  22.5	&  90	&  22.5	&  90	&  90\\
\hline
	  Trial2.1	&  90	&  67.5	&  45	&  22.5	&  90	&  112.5	&  90	&  45		&  90	&  67.5	&  315	&  22.5	&  90	&  67.5	&  90	&  90\\
\hline
	  Trial2.2	&  90	&  112.5	&  45	&  22.5	&  90	&  112.5	&  90	&  45		&  90	&  112.5	&  315	&  22.5	&  90	&  67.5	&  90	&  90\\
\hline
	  Trial2.3	&  90	&  67.5	&  315	&  22.5	&  90	&  112.5	&  90	&  45		&  90	&  67.5	&  45	&  22.5	&  90	&  67.5	&  90	&  90\\
\hline
	  Trial2.4	&  90	&  112.5	&  315	&  22.5	&  90	&  112.5	&  90	&  45		&  90	&  112.5	&  45	&  22.5	&  90	&  67.5	&  90	&  90\\
\hline
	  Trial2.5	&  90	&  67.5	&  45	&  22.5	&  90	&  67.5	&  90	&  45		&  90	&  67.5	&  315	&  22.5	&  90	&  22.5	&  90	&  90\\
\hline
	  Trial2.6	&  90	&  112.5	&  45	&  22.5	&  90	&  67.5	&  90	&  45		&  90	&  112.5	&  315	&  22.5	&  90	&  22.5	&  90	&  90\\
\hline
	  Trial2.7	&  90	&  67.5	&  315	&  22.5	&  90	&  67.5	&  90	&  45		&  90	&  67.5	&  45	&  22.5	&  90	&  22.5	&  90	&  90\\
\hline
	  Trial2.8	&  90	&  112.5	&  315	&  22.5	&  90	&  67.5	&  90	&  45		&  90	&  112.5	&  45	&  22.5	&  90	&  22.5	&  90	&  90\\
\hline																
	  Trial3.1	&  90	&  67.5	&  90	&  112.5	&  45	&  22.5	&  90	&  45		&  90	&  67.5	&  90	&  67.5	&  315	&  22.5	&  90	&  90\\
\hline
	  Trial3.2	&  90	&  112.5	&  90	&  112.5	&  45	&  22.5	&  90	&  45		&  90	&  112.5	&  90	&  67.5	&  315	&  22.5	&  90	&  90\\
\hline
	  Trial3.3	&  90	&  67.5	&  90	&  67.5	&  45	&  22.5	&  90	&  45		&  90	&  67.5	&  90	&  22.5	&  315	&  22.5	&  90	&  90\\
\hline
	  Trial3.4	&  90	&  112.5	&  90	&  67.5	&  45	&  22.5	&  90	&  45		&  90	&  112.5	&  90	&  22.5	&  315	&  22.5	&  90	&  90\\
\hline
	  Trial3.5	&  90	&  67.5	&  90	&  112.5	&  315	&  22.5	&  90	&  45		&  90	&  67.5	&  90	&  67.5	&  45	&  22.5	&  90	&  90\\
\hline
	  Trial3.6	&  90	&  112.5	&  90	&  112.5	&  315	&  22.5	&  90	&  45		&  90	&  112.5	&  90	&  67.5	&  45	&  22.5	&  90	&  90\\
\hline
	  Trial3.7	&  90	&  67.5	&  90	&  67.5	&  315	&  22.5	&  90	&  45		&  90	&  67.5	&  90	&  112.5	&  45	&  22.5	&  90	&  90\\
\hline
	  Trial3.8	&  90	&  112.5	&  90	&  67.5	&  315	&  22.5	&  90	&  45		&  90	&  112.5	&  90	&  112.5	&  45	&  22.5	&  90	&  90\\
\hline																		
	  Trial4.1	&  45	&  67.5	&  45	&  22.5	&  45	&  22.5	&  90	&  45		&  45	&  67.5	&  315	&  22.5	&  315	&  22.5	&  90	&  90\\
\hline
	  Trial4.2	&  135	&  112.5	&  45	&  22.5	&  45	&  22.5	&  90	&  45		&  135	&  112.5	&  315	&  22.5	&  315	&  22.5	&  90	&  90\\
\hline
	  Trial4.3	&  45	&  67.5	&  315	&  22.5	&  45	&  22.5	&  90	&  45		&  45	&  67.5	&  45	&  22.5	&  315	&  22.5	&  90	&  90\\
\hline
	  Trial4.4	&  135	&  112.5	&  315	&  22.5	&  45	&  22.5	&  90	&  45		&  135	&  112.5	&  45	&  22.5	&  315	&  22.5	&  90	&  90\\
\hline
	  Trial4.5	&  45	&  67.5	&  45	&  22.5	&  315	&  22.5	&  90	&  45		&  45	&  67.5	&  315	&  22.5	&  45	&  22.5	&  90	&  90\\
\hline
	  Trial4.6	&  135	&  112.5	&  45	&  22.5	&  315	&  22.5	&  90	&  45		&  135	&  112.5	&  315	&  22.5	&  45	&  22.5	&  90	&  90\\
\hline
	  Trial4.7	&  45	&  67.5	&  315	&  22.5	&  315	&  22.5	&  90	&  45		&  45	&  67.5	&  45	&  22.5	&  45	&  22.5	&  90	&  90\\
\hline
	  Trial4.8	&  135	&  112.5	&  315	&  22.5	&  315	&  22.5	&  90	&  45		&  135	&  112.5	&  45	&  22.5	&  45	&  22.5	&  90	&  90\\
\hline

\end{tabular}}
\caption{Angles of wave-plates}
\end{table}

\end{document}